\documentclass{elsart}
\usepackage[pctex32]{graphics}
\usepackage{graphicx,amsmath,verbatim,amssymb,psfrag}

\begin{document}
\begin{frontmatter}
\title{Efficiency characterization of a large neuronal network: a causal information approach}
\author[DtoFis,Iflysib]{Fernando Montani},
\ead{fmontani@gmail.com}
\corauth[cor]{Corresponding author}
\author[DtoFis,Iflysib]{Emilia B. Deleglise},
\author[Maceio,UBA]{Osvaldo A. Rosso}.

\address[DtoFis]{Departamento de F\'{\i}sica, Facultad de Ciencias Exactas, \\
           UNLP Calle 49 y 115. C.C. 67 (1900), La Plata, Argentina.}

\address[Iflysib]{IFLYSIB, Universidad Nacional de La Plata, La Plata, Argentina}

\address[UBA]{Laboratorio de Sistemas Complejos, Facultad de Ingenier\'{\i}a,\\
              Universidad de Buenos Aires (UBA).\\
              (1063) Av. Paseo Col\'on 840, Ciudad Aut\'onoma de Buenos Aires, Argentina.}

\address[Maceio]{Instituto de F\'{\i}sica,
              Universidade Federal de Alagoas (UFAL).\\
              BR 104 Norte km 97,
              57072-970 Macei\'o, Alagoas, Brazil.}
\begin{abstract}
When inhibitory neurons constitute about 40\% of neurons they could have an important antinociceptive role, as they
would easily regulate the level of activity of other neurons.
We consider a simple network of cortical spiking neurons with axonal conduction delays and spike timing dependent plasticity, representative of a cortical column or hypercolumn with large proportion of inhibitory neurons.
Each neuron fires following a Hodgkin-Huxley like dynamics and it is interconnected randomly to other neurons.
The network dynamics is investigated estimating Bandt and Pompe probability distribution function associated to the interspike intervals and taking different degrees of inter-connectivity across neurons.
More specifically we take into account the fine temporal ``structures'' of the complex neuronal signals not just by using the probability distributions associated to the inter spike intervals, but instead considering much more subtle measures accounting for their causal information: the Shannon permutation entropy, Fisher permutation information and permutation statistical complexity.
This allows us to investigate how the information of the system  might saturate to a finite value as
the degree of inter-connectivity across neurons grows, inferring the emergent dynamical properties of the system.
\newenvironment{keywords}{
       \list{}{\advance\topsep by0.35cm\relax\small
       \leftmargin=1cm
       \labelwidth=0.35cm
       \listparindent=0.35cm
       \itemindent\listparindent
       \rightmargin\leftmargin}\item[\hskip\labelsep
                                     \bfseries Keywords:]}
     {\endlist}
\end{abstract}
\maketitle
\end{frontmatter}

\section{Introduction}
\label{Sec:Intro}

The central assumption of theoretical neuroscience is that the brain computes. That is, in general, it is accepted
that the brain is a complex dynamical system whose state variables encode information about the outside world.  The computation it is therefore equal to the coding more dynamic. The detection of subtle changes in brain dynamics
it is therefore of importance to investigate the dynamic of functional interactions across neurons.

EEG and fMRI data are two important measures of brain activity. EEG reflects the brain's electrical activity such as post-synaptic potentials, and fMRI detects blood flow, motion and equilibrium under the action of external forces.
The action potential (or spike) is an explosion of electrical activity that is created by a depolarizing current.
Single-unit recordings measures spikes responses of a single neuron, and constitute an important methodology
for understanding mechanisms and functions within the cerebral cortex. The development of micro-electrode arrays allowed recording from multiple units at
the same time, and constitutes an important method for studying functional interactions among neuronal populations.
In the field of experimental
neurophysiology, one of the most common magnitude used for the study of the neuronal
network dynamics is the variance of the inter spike intervals (ISIs)
\cite{Anderson1994,Yadid2008,Lundstrom2006,Oswald2007}.

Detection of dynamic changes in complex systems is one of the most relevant issues in neuroscience.
Due to the occurrence of noise and artifacts in various forms, it is often not easy to get reliable information from a series of measurements. That is, brain activity measures are
characterized by a variety of dynamic variables which are noisy,
nonstationary, nonlinear and rife with temporal discontinuities.

Bandt and Pompe \cite{Bandt2002} have proposed a robust approach to time series analysis on the basis
of counting ordinal patterns by introducing the concept of permutation entropy for quantifying the complexity of a system behind a time series.
Thus, the ordinal structures of the time series instead of the values themselves are considered \cite{Zanin2012}.
This methodology has been applied for investigating EEG and FMRi signals
\cite{Schindler2011,Veisi2007,Cao2004,Ouyang2008,Bruzzo2008,Li2007,Olofsen2008,Jordan2008,Nicolau2012,
Robinson2013,Schroter2012,Rummel2013}.

The nonlinear dynamic effects of the spiking activity of in vivo neuronal networks
cannot be fully captured by simple calculations of ISIs. More specifically,
estimating the variance of the ISIs may not always provide reliable information about the system dynamics. Importantly, the application of the Bandt and Pompe methodology \cite{Bandt2002} to spiking neural data has not been widely investigated. In the current paper we considered the effective statistical complexity measure (SCM) introduced by Lamberti {\it et al.\/} \cite{Lamberti2004}, (also called MPR complexity) that allows us to detect essential details of the dynamics of a neural network.

One of the simplest computational models of neurons is the Simple Model of Spiking Neurons by Izhikevich \cite{Izhikevich2003,Izhikevich2004}. This model just uses a two-dimensional (2-D) system of ordinary differential equations and four parameters to generate several different types of neural dynamics.
In order to reproduce some of the dynamic properties of a population of neurons, we use a neural network capable of emulating several of the biological characteristics of
brain oscillation patterns and spontaneous brain activity \cite{Izhikevich2004,Izhikevich2006}.
We consider a population of neurons with large
number inhibitory neurons, as it helps to rapidly phase lock neural populations and induce synchronization at small time
windows and produces stable firing patterns \cite{Lytton1992}. In particular, large amounts of inhibitory neurons can have a crucial role in regulating the level of activity of other neurons \cite{Yuan2011}.
Thus, inhibitory neurons have an important antinociceptive role when they constitute about 40\% of the
population \cite{Tiong2011}.


The nervous system can extract only a finite amount of
information about sensory stimuli, and, in subsequent
stages of processing, the amount of information cannot
exceed the amount extracted and therefore information should saturate as the number of interconnected neurons
becomes large enough. This is, the correlations must behave in such way that information
should saturate, or reach a maximum, as the number of interconnected neurons increases \cite{Averbeck2006}.

In this paper we use the ISIs of a very simple spiking model \cite{Izhikevich2007} of a population of neurons to test the performance of
the variance of the ISIs in comparison with more
subtle measures accounting for the nonlinear dynamic effects of the temporal
signal: the Shannon entropy \cite{Rosso2009A,Rosso2009B}, the MPR statistical complexity \cite{Rosso2009A,Rosso2009B} and the Fisher information measure \cite{Ferri2009,Quiroga09}.
We show that simple estimations of the variance of ISIs do not capture the saturation properties of the information as the number
 of interconnected neurons increases.
Our proposal therefore is to use the Bandt-Pompe permutation methodology
for the evaluation of probability distribution function
(PDF) associated with a time series \cite{Bandt2002} in order to  characterize the dynamics of the spiking neural activity when simulating a cortical network.
By estimating Fisher information versus MPR statistical complexity/Shannon entropy \cite{Rosso2010,Olivares2012A,Olivares2012B} of the ISIs signals, we show
that it is possible to quantify the optimal amount of interconnected neurons that maximizes information transmission
within a simple model that we use as ``representative of a cortical hypercolumn with large proportion of inhibitory neurons''.

Based on the quantification of the ordinal ``structures'' present in the ISIs and their local influence on the associated probability distribution, we incorporate the time series own temporal causality through a system of easy algorithm implementation and computation.
We show that estimating the variance of the ISIs does not help to understand
the dynamics of the system, and thus statistical measures
taking into account the time causality of the signal are needed. Our approach allows us to estimate the saturation properties of the neuronal network,
quantifying the causality of the signal, and inferring the emergent dynamical properties
of the system as the number of interconnected neurons increases.


\section{Neuronal network simulation model}
\label{Sec:Modelo}

Neurons fire spikes when they are near a bifurcation from resting to spiking activity, and it is the delicate
balance between noise, dynamic currents and initial condition what determines the phase diagram of neural activity.
While there are a huge number of possible ionic mechanisms of excitability and spike generation,
there are just four bifurcation mechanisms that can result in such a transition.
That is, there are many ionic mechanisms of spike generation, but only four
generic bifurcations of equilibrium. These bifurcations divide neurons into four categories: integrators
or resonators, monostable or bistable \cite{Izhikevich2007}.

Bifurcation methodologies \cite{Izhikevich2007} allow us to accurately reproduce the biophysical properties
of Hodgkin-Huxley neuronal models by just taking a two-dimensional
system of ordinary differential equations and four different parameters. This model is named
{\it simple model of spiking neurons\/}
\cite{Izhikevich2003}, and the system of ordinary differential equations reads:
\begin{equation}
{{dv}\over {dt}} ~=~0.04~ v^2 + 5~v + 140~ u + I
\label{izhi1}
\end{equation}
\begin{equation}
{{du}\over {dt}} ~=~a~(b~v - u)
\label{izhi2}
\end{equation}
with the auxiliary after-spike resetting
\begin{equation}
{\rm if} \  v \geq +30~mV, \quad {\rm then} \quad
\begin{cases}
v \leftarrow c \\
u \leftarrow u + d. \\
\end{cases}
\label{izhi}
\end{equation}
Where $v$ is the membrane potential of the neuron, and $u$
is a membrane recovery variable, which accounts for the activation
of $K+$ ionic currents and inactivation of $Na+$ ionic currents, and gives
negative feedback to $v$. Thus we are just considering a two-dimensional
(2-D) system of ordinary differential equations of two variables $u$ and $v$,
and all the known types of neurons can be reproduced by taking different
values of the four parameters $a$, $b$, $c$ and $d$.
After the spike reaches its apex at $+30~mV$ (not to be confused with the firing threshold)
the membrane voltage and the recovery variable are reset according to Eq.~(\ref{izhi}).
The variable $I$ accounts for the inputs to the neurons \cite{Izhikevich2003}.

Summarizing, each neuron can be described by a simple spiking model that allows us to reproduce several of the
most fundamental neurocomputational features of biological neurons \cite{Izhikevich2004}.
Here, we consider a network simulation model in which the number for interconnected neurons is a parameter
under control,  and each neuron can be interconnected randomly with two or more neurons \cite{Izhikevich2003, Izhikevich2004,Izhikevich2006}.
The inhibitory inputs hyperpolarize the potential and move it
away from the threshold. In contrast, excitatory inputs depolarize the membrane potential (i.e., they bring it closer to the ``firing threshold'').

Regular spiking (RS) \cite{Izhikevich2003} neurons are the major class of excitatory neurons ($b > 0$). RS neurons are the most typical neurons in the cortex.
In our current simple model for the excitatory neurons we take $a=0.02$ and $b=0.2$ with fixed values; while $c=-65+15\cdot(r_{de})^2$ and $d=8-6\cdot(r_{de})^2$, where $r_{de}$ is random vector (between zero and one) of the size of the number of excitatory neurons.

Fast spiking (FS) neurons are the ones that fire periodic trains of action potentials without
any adaptation. FS neurons are inhibitory neurons that have $a=0.1$ and $b=0.25-0.05 \cdot r_{di}$, where $r_{di}$ is a random vector (between zero and one) of the size of the number of inhibitory neurons. Furthermore $c=-65~mV$ and $d =2$.
Positive synaptic weights are taken for excitatory neurons and negative, for the inhibitory ones. This simple model reproduces some biologically plausible phenomena such as patterns of spontaneous cortical activity including brain wave like oscillations.

The network simulation takes into account cortical spiking neurons with axonal conduction delays and spike-timing-dependent plasticity (STDP). The magnitude of the synaptic weight between pre- and postsynaptic neurons depends on the timing of the spikes according to the STDP rules. That is, the weight of the synaptic connection from the pre- to postsynaptic neurons grows as  $0.12 \cdot \exp{(t / t_0)}$ if the postneuron fires after the presynaptic spike, and if the order is reversed, it decreases as $0.1 \cdot \exp{(-t / t_0)}$, where $t_0 = 20~ms$ \cite{Izhikevich2006}.
Importantly, the interplay between conduction delays and STDP helps the spiking neurons to produce stable firing patterns that would not be possible without these assumptions. Each neuron in the network is described by the
simple model of spiking neurons of \cite{Izhikevich2003}, which has been described above in Eq.~(\ref{izhi1}) to Eq.~(\ref{izhi}).

In this simple model we choose to account for the STDP as its interplay with conduction delays
that
helps the spiking neurons to spontaneously self-organize into groups with patterns of time-locked activity.
Thus, the STPD produces stable firing patterns emulating spontaneous brain activity that would not be possible without this
assumption. Although the patterns are random, reflecting connectivity within the cortex, one could implement sophisticated anatomy and the two sparse networks shown in this section are representative of a
cortical hypercolumn \cite{Izhikevich2006}.

Since we cannot simulate an infinite-dimensional system on a finite-dimensional lattice, we choose a large network with
a finite-dimensional approximation taking a time resolution of $1~ms$. The main idea of our methodology
is to use this simple model to emulate the spontaneous activity when a large number of neurons
are taken in order to show the advantage of using causal quantifiers when considering the ISIs.

To illustrate the simple model that we are using,
{\it (a) \/} as a first network case we consider
a network that consists of $n = 1000$ neurons, with $n_e = 550$ being of excitatory regular spiking (RS) type,
and the remaining $n_i = 450$ of inhibitory fast spiking (FS) type ($n = n_e + n_i$).
{\it (b) \/} As a second network, we take a network that consists of $n = 900$ neurons, with $n_e = 500$ being of excitatory (RS) type,
and the remaining $n_i = 400$ of inhibitory (FS) type ($n=n_e+n_i$);
{\it (c) \/} Finally, as third network, we consider,
 a network that consists of $n = 800$ neurons, with $n_e = 450$ being of excitatory (RS) type,
and the remaining $n_i = 450$ of inhibitory (FS) type ($n=n_e+n_i$).
In all the cases considered, each excitatory regular spiking type neuron, as well as each inhibitory fast spiking neuron,
is connected to $m=2$, $4$, $6$, $8$, $10$, $20$, $30$, $40$, $60$, $80$, $100$ and $120$ random neurons, so that the probability of
connection is $m/n$.
In the case of inhibitory FS neurons, the connections are with excitatory neurons only.
In all cases synaptic connections among neurons have fixed conduction delays, which are random integers between $1~ms$ and $10~ms$.

In the next sections we show that understanding how neural information saturates as the number of neurons increases requires the development of an appropriate mathematical framework accounting for the ordinal ``structures'' present in the time series.
We will also show how to quantify the causal information of the ISIs.
For the sake of simplicity in this paper we will study a simple row signal of spontaneous neural activity.
We will investigate the effect of increasing the network connectivity of a simulated cortical hypercolumn, with
large percentage of inhibitory neurons,
by quantifying the degree of correlation with and without considering the causality information present in the ISIs.

\section{Information Theory quantifiers}
\label{Sec:Information_Therory}

\subsection{Shannon Entropy, Fisher Information Measure and  MPR Statistical Complexity}
\label{Sec:Info-quantifiers}

Sequences of measurements (or observations) constitute the basic elements for the study of natural phenomena.
In particular, from these sequences, commonly called time series, one should judiciously extract information on the dynamical systems under study.
We can define an Information Theory quantifier as a measure that is able to characterize some property of the probability distribution function associated with these time series of a given row signal (i.e., ISIs).
Entropy, regarded as a measure of uncertainty, is the most paradigmatic example of these quantifiers.

Given a time series ${\mathcal X}(t) \equiv \{ x_t ; t = 1, \cdots , M \}$, a set of $M$ measures of the observable ${\mathcal X}$ and
the associated PDF, given by $P  \equiv \{ p_j ; j = 1, \cdots , N \}$ with $\sum_{j=1}^{N} p_j = 1$  and $N$ the number of possible states of
the system under study, the Shannon's logarithmic information measure \cite{Shannon1949} is defined by
\begin{equation}
\label{entropy}
S[P] = - \sum_{j=1}^{N} p_j \ln( p_j ) \ .
\end{equation}
This functional is equal to zero when we are able to predict with full certainty which of the possible outcomes $j$, whose probabilities are
given by $p_j$, will actually take place.
Our knowledge of the underlying process, described by the probability distribution, is maximal in this instance.
In contrast, this knowledge is commonly minimal for a uniform distribution $P_e = \{ p_j = 1/N, \forall j = 1, \cdots, N \}$.

The Shannon entropy $S$ is a measure of ``global character" that is not too sensitive to strong changes in the PDF taking place in
small region.
Such is not the case with the Fisher information measure \cite{Fisher_1922,Frieden2004}
\begin{equation}
\label{fisher}
F[f]~=~\int    { {|\vec \nabla f(x)|^2}  \over {f(x)} }  ~dx \  ,
\end{equation}
which constitutes a measure of the gradient content of the distribution $f$ (continuous PDF), thus being quite sensitive even to tiny localized
perturbations.

The Fisher information measure can be variously interpreted as a measure of the ability to estimate a parameter, as the amount of information
that can be extracted from a set of measurements, and also as a measure of the state of disorder of a system or phenomenon
\cite{Frieden2004,Mayer2006}, its most important property being the so-called Cramer-Rao bound.
It is important to remark  that the gradient operator significantly influences the contribution of minute local $f$-variations to the Fisher information
value, so that the quantifier is called a ``local" one.
Note that  Shannon entropy decreases with skewed distribution, while  Fisher information increases in such a case.
Local sensitivity is useful in scenarios whose description necessitates an appeal to a notion of ``order" \cite{Rosso2010,Olivares2012A,Olivares2012B}.
The concomitant problem of loss of information due to the discretization has been thoroughly studied (see, for instance,
\cite{Zografos1986,Pardo1994,Madiman2007} and references therein) and, in particular, it entails the loss of Fisher's shift-invariance,
which is of no importance for our present purposes.

Let us now consider the activity of a population of neurons represented by an $\aleph$-dimensional vector
$x_{\aleph} =(x_1;~x_2;~ \cdots;~x_{\aleph})$ where the randomness of the spiking activity produces an inherent accuracy
due to the trial to trial fluctuations. Fisher information $F$ provides a measure of the encoding
accuracy
\begin{eqnarray}
\frac{1}{F}~=~ \sum_{i=1}^{\aleph} \left( \left< \left[\frac{ \partial \ln P(n_{sp}| {x}, {\Delta T})}{ \partial x_i} \right]^2 \right> \right)^{-1}
\end{eqnarray}
where its inverse is the Cramer-Rao lower bound $E[\epsilon^2] \geqq \frac{1}{F}$ \cite{Brunel98, Sejnowski99, Quiroga09}.
Note that $\epsilon=\epsilon_1+ \ldots + \epsilon_{\aleph} $ is the square error in a single trial, and $n_{sp}$
is the number of spikes.
Considering the general case in which the probability distribution is a function of
the mean firing rate and the time windows, then these conditions are sufficient to show that \cite{Sejnowski99}:

\begin{eqnarray}
{F} ~ =~  \; \eta \; \sigma^{\aleph-2} K_{\phi}(f,\Delta T,\aleph)
\label{eq:Fhis}
\end{eqnarray}
where $\sigma$ is the variance (or tuning width that can take any positive real value) and $K_{\phi}(f,\Delta T,\aleph)$ denotes the dependence
on the $\aleph$-dimensional space of the encoded variable, and $\eta$ is a density factor proportional to the number
of inactive neurons.

Fisher information should saturate as the number of neurons, or the network
interconnectivity $m$, becomes very large \cite{Averbeck2006}. Note that one
would expect, therefore, that the variance should not diverge as the interconnectivity across
neurons increases. That is, the variance should reach a maximum for some given value of $m$ as
the interconnectivity becomes higher.

For Fisher information measure computation (discrete PDF) we follow the proposal of Ferri and coworkers \cite{Ferri2009} (among others)
\begin{equation}
\label{fisher-disc}
F[P]~={1\over 4}~\sum_{i=1}^{N-1}~2 { {(p_{i+1} - p_i)^2}  \over {( p_{i+1} + p_i)} } \ .
\end{equation}
If our system is in a very ordered state and thus is represented by a very narrow PDF, we have a Shannon entropy $S \sim 0$ and a
Fisher information measure $F \sim F_{max}$.
On the other hand, when the system under study lies  in a very disordered state one gets an almost flat PDF and $S \sim S_{max}$, while
$F \sim 0$.
Of course,  $S_{max}$ and $F_{max}$ are, respectively, the maximum values for the Shannon entropy and Fisher information measure.
One can state that the general behavior of the Fisher information measure is opposite to that of the Shannon entropy \cite{Pennini2005}.

It is well known, however, that the ordinal structures present in a process is not quantified by randomness measures and,
consequently, measures of statistical or structural complexity are necessary for a better understanding (characterization) of the system dynamics
represented by their time series \cite{Feldman1998}.
The opposite extremes of perfect order (i.e., a periodic sequence) and maximal randomness (i.e., a fair coin toss) are very simple
to describe because they do not have any structure.
The complexity should be zero in these cases.
At a given distance from these extremes, a wide range of possible ordinal structures exists.
The complexity measure allows one to quantify this array of behavior \cite{Feldman2008}.
We consider the MPR complexity \cite{Lamberti2004} as it is able quantify critical details of dynamical processes underlying the data set.

Based on the seminal notion advanced by L\'opez-Ruiz {\it et al.\/} \cite{LMC1995}, this statistical complexity measure is defined through
the product
\begin{equation}
{C}_{JS}[P]~=~{\mathcal Q}_{J}[P,P_e] \cdot {H}_S[P]
\label{complexity}
\end{equation}
of the normalized Shannon entropy
\begin{equation}
\label{normalized-entropy}
{H}_S[P]~=~ S[P] / S_{max}
\end{equation}
with $S_{max} = S[ P_e ] = \ln{ N }$, ($0 \leq {\mathcal H}_S \leq 1$) and the disequilibrium ${\mathcal Q}_{J}$ defined in terms of the
Jensen-Shannon divergence.
That is,
\begin{equation}
\label{disequilibrium}
{\mathcal Q}_{J} [ P, P_e] ~= Q_{0}~{\mathcal J}[ P, P_e]
\end{equation}
with
\begin{equation}
\label{Jensen}
{\mathcal J} [ P, P_e]~=~S[(P + P_e)/2 ] - S[ P ]/2 - S[P_e]/2
\end{equation}
the above-mentioned Jensen-Shannon divergence and $Q_0$, a normalization constant ($0 \leq {\mathcal Q}_{J} \leq 1$), are equal to the inverse
of the maximum possible value of ${\mathcal J} [P,P_e]$.
This value is obtained when one of the components of $P$, say $p_m$, is equal to one and the remaining $p_j$ are equal to zero.
The Jensen-Shannon divergence, which quantifies the difference between two (or more) probability distributions, is especially
useful to compare the symbolic composition between different sequences \cite{Grosse2002}.
Note that the above introduced SCM depends on two different probability distributions, the one associated with the system under
analysis, $P$, and the uniform distribution, $P_e$.
Furthermore, it was shown that for a given value of ${H}_{S}$, the range of possible $C_{JS}$ values varies between a
minimum ${C}_{min}$ and a maximum ${C}_{max}$, restricting the possible values of the SCM in a given
complexity-entropy plane \cite{Martin2006}.
Thus, it is clear that important additional information related to the correlational structure between the components of the physical
system is provided by evaluating the statistical complexity measure.

\subsection{The Bandt-Pompe approach to the PDF determination}
\label{Sec:band-Pompe}

The study and characterization of time series ${\mathcal X}(t)$ by recourse to Information Theory tools assume that the underlying
PDF is given a priori.
In contrast, part of the concomitant analysis involves extracting the PDF from the data and there is no univocal procedure with which
everyone agrees.
Almost ten years ago Bandt and Pompe (BP) introduced a successful methodology for the evaluation of the PDF associated with scalar time
series data using a symbolization technique \cite{Bandt2002}.
For a didactic description of the approach, as well as, its main biomedical and econophysics applications, see \cite{Zanin2012}.

The pertinent symbolic data are {\it (i)\/} created by ranking the values of the series and {\it (ii)\/} defined by reordering the
embedded data in ascending order, which is tantamount to a phase space reconstruction with embedding dimension (pattern length) $D$
and time lag $\tau$. 
In this way it is  possible to quantify the diversity of the ordering symbols (patterns) derived from a scalar time series.
Note that the appropriate symbol sequence arises naturally from the time series and no model-based assumptions are needed.
In fact, the necessary ``partitions" are devised by comparing the order of neighboring relative values rather than by apportioning
amplitudes according to different levels.
This technique, as opposed to most of those in current practice, takes into account the temporal structure of the time series generated
by the physical process under study.
This feature allows us to uncover important details concerning the ordinal structure of the time series
\cite{Olivares2012B,Rosso2007,Rosso2013}
and can also yield information about temporal correlation \cite{Rosso2009A,Rosso2009B}.
It is clear that this type of analysis of time series entails losing some details of the original series' amplitude information.
Nevertheless, by just referring to the series' intrinsic structure, a meaningful difficulty reduction has indeed been achieved by Bandt and
Pompe with regard to the description of complex systems.
The symbolic representation of time series by recourse to a comparison of consecutive ($\tau = 1$) or nonconsecutive ($\tau > 1$) values
allows for an accurate empirical reconstruction of the underlying phase-space, even in the presence of weak (observational and dynamic)
noise \cite{Bandt2002}.
Furthermore, the ordinal patterns associated with the PDF is invariant with respect to nonlinear monotonous transformations.
Accordingly, nonlinear drifts or scaling artificially introduced by a measurement device will not modify the estimation of quantifiers,
a nice property if one deals with experimental data (see, e.g., \cite{Saco2010}).
These advantages make the Bandt and Pompe methodology more convenient than conventional methods based on range partitioning
(i.e., PDF based on histograms).

Additional advantages of the method reside in
{\it (i)\/} its simplicity, we need few parameters: the pattern length/embedding dimension $D$ and the embedding delay $\tau$, and
{\it (ii)\/} the extremely fast nature of the pertinent calculation process \cite{Keller2005}.
The BP methodology can be applied not only to time series representative of low dimensional dynamical systems, but also to any type
of time series (regular, chaotic, noisy, or reality based).
In fact, the existence of an attractor in the $D$-dimensional phase space is not assumed.
The only condition for the applicability of the Bandt-Pompe methodology is a very weak stationary assumption (that is, for $k \leq D$,
the probability for $x_t < x_{t+k}$ should not depend on $t$ \cite{Bandt2002}).

To use the Bandt and Pompe \cite{Bandt2002} methodology for evaluating the PDF, $P$, associated with the time series (dynamical system)
under study, one starts by considering partitions of the pertinent $D$-dimensional space that will hopefully ``reveal" relevant details
of the ordinal structure of a given one-dimensional time series ${\mathcal X}(t) = \{ x_t; t= 1, \cdots, M\}$ with embedding dimension
$D > 1$ ($D \in {\mathbb N}$) and embedding time delay $\tau$ ($\tau \in {\mathbb N}$).
We are interested in ``ordinal patterns" of order (length) $D$ generated by
$(s)~\mapsto~ \left(~x_{s-(D-1)\tau},~x_{s-(D-2)\tau},~\cdots, \\ ~x_{s-\tau},~x_{s}~\right)$, which assigns to each time $s$ the $D$-dimensional
vector of values at times $s, s-\tau,\cdots,s-(D-1)\tau$.
Clearly, the greater the $D-$value, is the more information on the past  is incorporated into our vectors.
By ``ordinal pattern" related to the time $(s)$ we mean the permutation $\pi=(r_0,r_1, \cdots,r_{D-1})$ of $[0,1,\cdots,D-1]$
defined by $x_{s-r_{D-1}\tau}~\le~x_{s-r_{D-2}\tau}~\le~\cdots~\le~x_{s-r_{1}\tau}~\le~x_{s-r_0\tau}$.
In order to get a unique result we set $r_i < r_{i-1}$ if $x_{s-r_{i}} = x_{s-r_{i-1}}$.
This is justified if the values of $x_t$ have a continuous distribution so that equal values are very unusual.
Thus, for all the $D!$ possible permutations $\pi$ of order $D$, their associated relative frequencies can be naturally computed
by the number of times this particular order sequence is found in the time series divided by the total number of sequences.

Consequently, it is possible to quantify the diversity of the ordering symbols (patterns of length $D$) derived from a scalar time series, by evaluating the so-called permutation entropy, the permutation statistical complexity and Fisher permutation information
measure.
Of course, the embedding dimension $D$ plays an important role in the evaluation of the appropriate probability distribution because $D$
determines the number of accessible states $D!$ and also conditions the minimum acceptable length
$M \gg D!$ of the time series that one needs in order to work with reliable statistics \cite{Rosso2007}.

Regarding to the selection of the other parameters, Bandt and Pompe suggested working with $4 \leq D \leq 6$ and specifically considered
an embedding delay $\tau = 1$ in their cornerstone paper \cite{Bandt2002}.
Nevertheless, it is clear that other values of $\tau$ could provide additional information.
It has been recently shown that this parameter is strongly related, if it is relevant, to the intrinsic time scales of the system
under analysis \cite{Zunino2010B,Soriano2011,Zunino2012}.

The Bandt and Pompe proposal for associating probability distributions to time series (of an underlying symbolic
nature), constitutes a significant advance in the study of non linear dynamical systems \cite{Bandt2002}.
The method provides univocal prescription for ordinary, global entropic quantifiers of the Shannon-kind.
However, as was shown by Rosso and coworkers \cite{Olivares2012A,Olivares2012B}, ambiguities arise in applying
the Bandt and Pompe technique with reference to the permutation of ordinal patterns. This happens if one wishes to employ the
BP-probability density to construct local entropic quantifiers, like Fisher information measure, that would
characterize
time series
generated by nonlinear dynamical systems.
The local sensitivity of Fisher information measure for discrete-PDFs is reflected in the fact that the specific
``$i$-ordering" of the discrete values $p_i$ must be seriously taken into account in evaluating the sum in
Eq. (\ref{fisher-disc}).
The pertinent numerator can be regarded as a kind of ``distance" between two contiguous probabilities.
Thus, a different ordering of the pertinent summands would lead to a different Fisher information value.
In fact, if we have a discrete PDF given by $P = \{ p_i, i = 1, \cdots , N\}$ we will have $N!$ possibilities.

The question is, which is the arrangement that one could regard as the ``proper" ordering?
The answer is straightforward in some cases, histogram-based PDF constituting a conspicuous example.
For such a procedure one first divides the interval $[a, b]$ (with $a$ and $b$ the minimum and
maximum values in the time series) into a finite number on nonoverlapping sub-intervals (bins).
Thus, the division procedure of the interval $[a, b]$ provides the natural order-sequence for the evaluation
of the PDF gradient involved in Fisher information measure.

\subsection{Causal information planes}
\label{Sec:Planos}

The above is based on Information Theory quantifiers ${H}_S$, ${F}$ and ${C}_{JS}$ evaluated using
Bandt and Pompe's PDF allow us to define three causality information planes: ${H} \times {C}$, ${H} \times {F}$
and ${C} \times {F}$.
The causality Shannon--complexity plane, ${H} \times {C}$, is based only on global characteristics of the associated
time series Bandt and Pompe PDF (both quantities are defined in terms of Shannon entropies), while the causality Shannon-Fisher plane,
${H} \times {F}$, and the causality complexity--Fisher,  ${C} \times {F}$ , are based on global and local
characteristics of the PDF.
In the case of  ${H} \times {C}$ the variation range is $[0, 1] \times [{C}_{min}, {C}_{max}]$
(with ${C}_{min}$ and ${C}_{max}$ the minimum and maximum statistical complexity values, respectively, for a given
${H}_{S}$ value \cite{Martin2006}), while in the causality planes  ${H} \times {F}$ and ${C} \times {F}$
the range is $[0, 1]\times [0, 1]$ in both cases.
These causal information planes have been profitably used to
separate and differentiate amongst chaotic and deterministic systems \cite{Rosso2007,Olivares2012B};
visualization and characterization of different dynamical regimes when the system parameters vary  \cite{Rosso2010,Olivares2012A,Olivares2012B};
time dynamic evolution \cite{Kowalski2007};
identifying periodicities in natural time series \cite{Bandt2005};
identification of deterministic dynamics contaminated with noise \cite{Rosso2012A,Rosso2012B}  and;
estimating intrinsic time scales of delayed systems \cite{Zunino2010B,Soriano2011,Zunino2012};
among other applications (see \cite{Zanin2012} and references therein).

\section{Results and Discussion}
\label{Sec:Results}

Cerebral cortex contains many inhibitory neurons of distinct types.
Understanding their connectivity and dynamics within the cortex is
essential to gain more knowledge of cortical information processing. Recent studies emphasize the essential role inhibitory neurons have in the development of proper circuitry in the cortex \cite{Tiong2011}. In particular, when inhibitory interneurons constitute about 40\% of neurons they have an important antinociceptive role, and are much more likely to
have a role in regulating the level of activity of other
neurons \cite{Tiong2011}.
We choose a network with large proportion of inhibitory neurons (about 40\% of the neurons)
as this helps the spiking neurons to spontaneously self-organize into groups with patterns of time-locked activity, producing stable firing patterns.  The objective is to use an information-theory-based approach to investigate the ordinal ``structures'' present in the time series of the ISIs of a model that emulates some of the characteristics of a cortical hypercolumn, considering a large
proportion of inhibitory neurons.

\begin{figure}

\begin{center}
\includegraphics[width=0.8\textwidth]{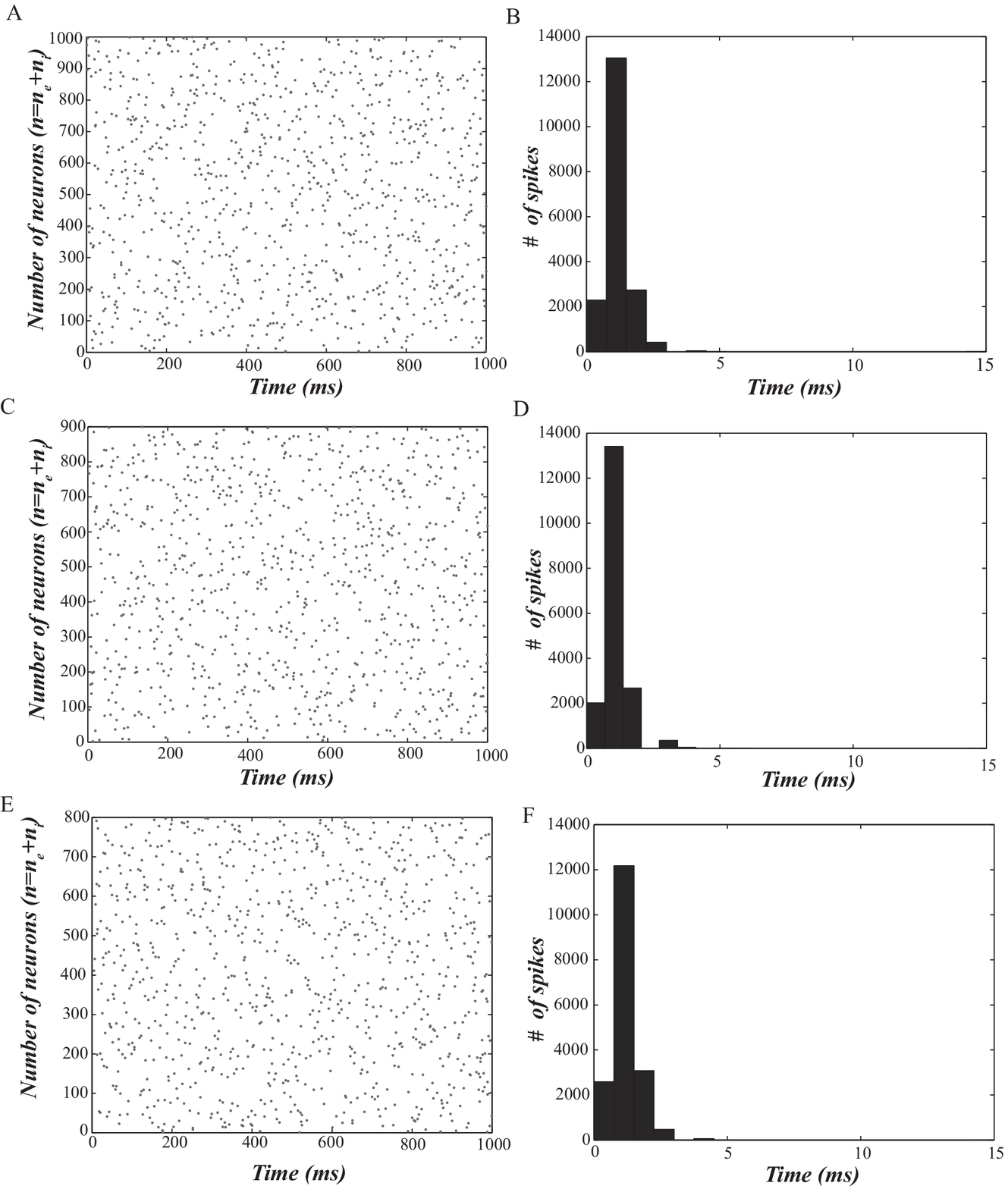}
\end{center}
\caption{\small{
\textbf{\emph{A})} Spike rasters, $n=1000$, $n_e=550$, $n_i=450$, $m=40$, taken in a time windows of $100~ms$.
\textbf{\emph{B})} Histogram of the interspike intervals (ISI) for $n=1000$, $n_e=550$, $n_i=450$ and $m=40$.
\textbf{\emph{C})} Same as in \emph{A} but considering $n=900$,  $n_e=500$, $n_i=400$ and $m=30$.
\textbf{\emph{D})} Same as in \emph{B} but considering $n=900$,  $n_e=500$, $n_i=400$ and $m=30$.
\textbf{\emph{E})} Same as in \emph{A} but considering $n=800$,  $n_e=450$, $n_i=350$ and $m=20$.
\textbf{\emph{F})} Same as in \emph{B} but considering $n=800$,  $n_e=450$, $n_i=350$ and $m=20$.}
\label{figu00} }
\end{figure}

In order to illustrate our method we run the simulation for $20000~ms$ taking a time resolution of $1~ms$ (time
windows $\Delta T = 1~ms$), which is enough to guarantee the condition $M \gg D!$ will be satisfied.
Figures~\ref{figu00} A, C and E show the spike rasters considered for the total number of neurons
$N=1000,~900$, and $800$ respectively, for a time window of $1000~ms$.
Figures~\ref{figu00} B, D and F show the interspike intervals considering neuronal total numbers
$N=1000, ~900$   and $800$ respectively, for the entire time we run the simulation.
Note that the previous plots only confirm that the neurons within the model fire with millisecond precision and we cannot gain
any significant information about the dynamics.

\begin{figure}

\begin{center}
\includegraphics[width=0.8\textwidth]{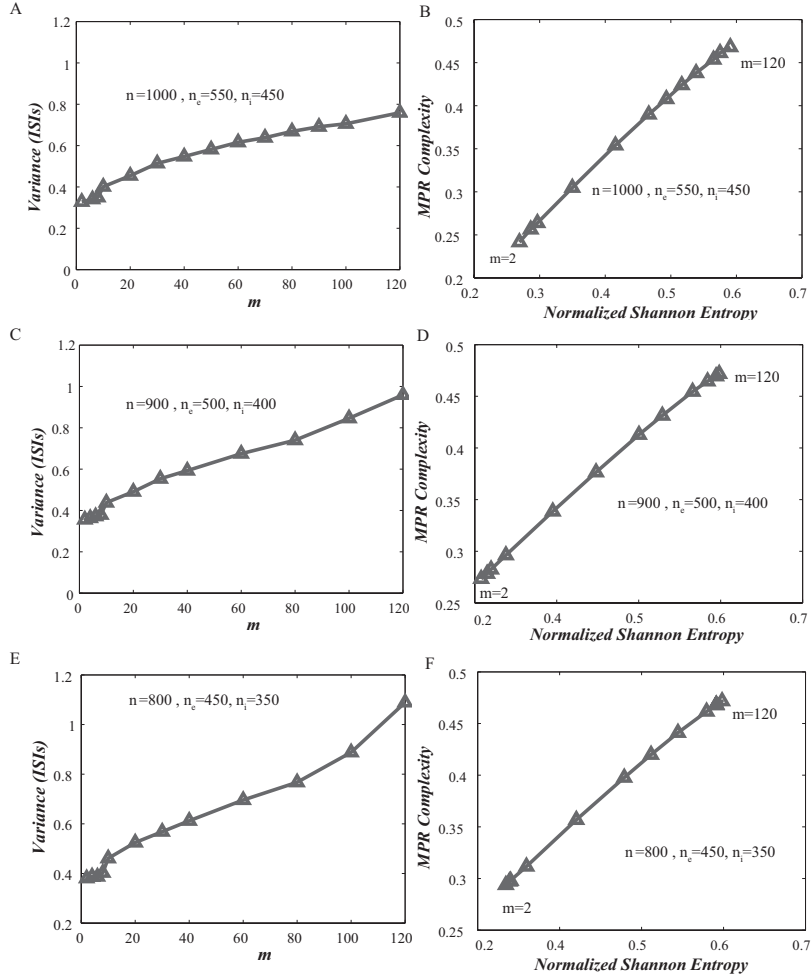}
\end{center}
\caption{\small{
\textbf{\emph{A})} Variance of the ISIs versus the number of interconnected neurons $m$ considering $n=1000$, $n_e=550$, $n_i=450$.
\textbf{\emph{B})} Causal MPR complexity versus normalized Shannon entropy ($H \times C$ plane), embedded dimensionality $D=6$. $n=1000$, $n_e=550$, $n_i=450$.
\textbf{\emph{C})} Same as in \emph{A} but considering $n=900$, $n_e=500$, $n_i=400$.
\textbf{\emph{D})} Same as in \emph{B} but considering $n=900$, $n_e=500$, $n_i=400$.
\textbf{\emph{E})} Same as in \emph{A} but considering $n=800$, $n_e=450$, $n_i=350$.
\textbf{\emph{F})} Same as in \emph{B} but considering $n=800$, $n_e=450$, $n_i=350$.
The grey triangles correspond to the different interconnectivities $m=2$, $4$, $6$, $8$, $10$, $20$, $30$, $40$, $60$, $80$, $100$ and $120$.}
\label{figu0} }
\end{figure}

Figures~\ref{figu0} A, C, and E show the variance $Rp=\sqrt{Var{(t_p)}}/<t_p>$ of the ISIs for three different simulated systems.
Note that an increasing degree of interconnectivity $m$ implies greater data dispersion.
Thus, this kind of analysis suggests that information does not saturate as the degree of interconnectivy
becomes higher. More specifically Eq.~(\ref{eq:Fhis}) implies that Fisher information
increases as the network connectivity becomes higher when considering a large population
of neurons of a cortical hypercolumn with large proportion of inhibitory neurons. This analysis performed through a classical tool
not accounting for the ``structures'' present in ISIs is far from the expected neurophysiologic behavior, which
suggests that Fisher information should reach a maximum as the number of interconnected neurons grows \cite{Averbeck2006}. The expected neurophysiological result is a consequence of the noise correlation, which causes
the amount of information to have a maximum as the number of interconnected becomes higher \cite{Averbeck2006, Shamir2001}.

In this section we use the Bandt and Pompe \cite{Bandt2002} methodology for evaluating the PDF, $P$, associated with the time series, considering an embedded dimensionality $D=6$ (with $\tau = 1$). This embedded dimensionality is enough to
efficiently capture causality information of the ordinal structure of the time series \cite{Bandt2002}.
Figures ~\ref{figu0} B, D and F show the informational causal plane of entropy versus complexity, $H \times C$. Note that the MPR statistical complexity grows
as the normalized entropy becomes higher and not much information can be gained about the system dynamics.

\begin{figure}

\begin{center}
 \includegraphics[width=0.8\textwidth]{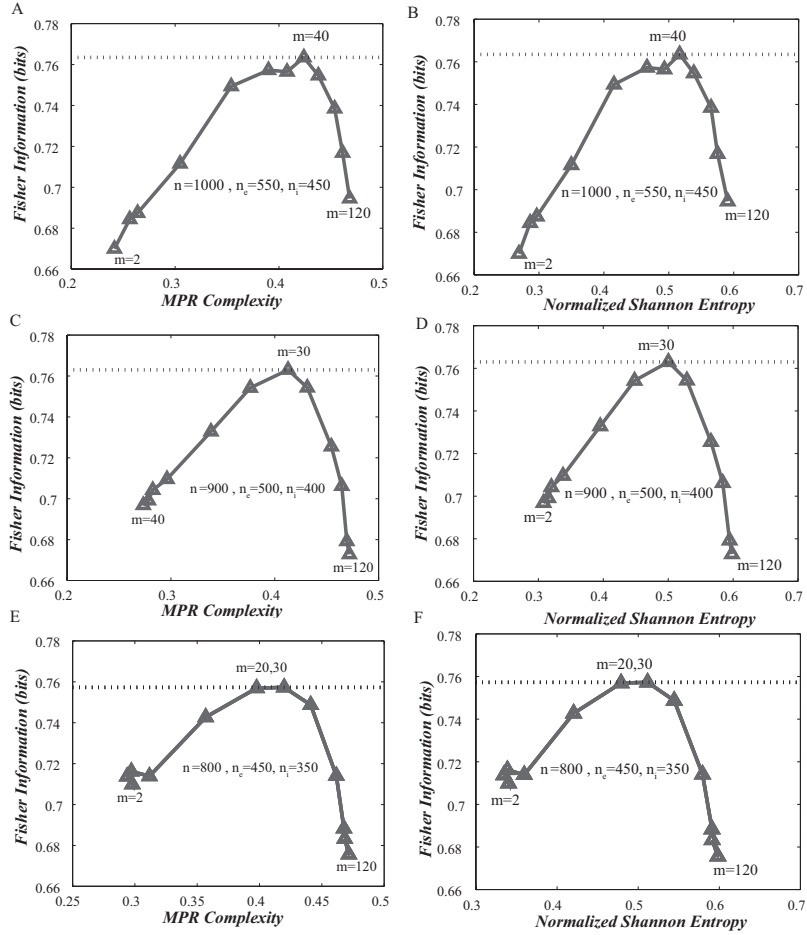}
 \end{center}
\caption{\small{
\textbf{\emph{A})} Causal Fisher information versus MPR complexity ($F \times C$ plane), embedded dimensionality $D=6$, $n=1000$, $n_e=550$, $n_i=450$. The maximum value occurs at $m=40$.
\textbf{\emph{B})} Causal Fisher information versus normalized Shannon permutation entropy ($F \times H$ plane), embedded dimension $D=6$, $n=1000$, $n_e=550$, $n_i=450$. The maximum value occurs at $m=40$.
\textbf{\emph{C})} Same as in \emph{A} but considering $n=900$, $n_e=500$, $n_i=400$.
The maximum value occurs at $m=30$.
\textbf{\emph{D})} Same as in \emph{B} but considering  $n=900$, $n_e=500$, $n_i=400$.
The maximum value occurs at $m=30$.
\textbf{\emph{E})} Same as in \emph{A} but considering  $n=800$, $n_e=450$, $n_i=350$.
The maximum values occur at $m= 20$ and $30$.
\textbf{\emph{F})} Same as in \emph{B} but considering  $n=800$, $n_e=450$, $n_i=350$.
The maximum values occur at $m= 20$ and $30$.
The grey triangles correspond to the different interconnectivities $m=2$, $4$, $6$, $8$, $10$,
$20$, $30$, $40$, $60$, $80$, $100$ and $120$.
When $m$ is raised, MPR complexity and entropy increase.}
\label{figu1} }
\end{figure}

Figures ~\ref{figu1} A, C and E show Fisher permutation information, Eq.(\ref{fisher-disc}), versus the permutation MPR statistical complexity
Eq.(\ref{disequilibrium}),
i.e., the causal information plane $F \times C$, for the same biophysical parameters used above.
Notice that Figures~\ref{figu1} A, C and E present a maximum at $m=40$, $m=30$ and $m=20,30$, respectively, for $n=1000$, $n=900$ and $n=800$.
A similar behavior can be observed in Figures~\ref{figu1} B, D and F, which show Fisher permutation information, Eq.(\ref{fisher-disc}), versus
permutation entropy, Eq.(\ref{normalized-entropy}), i.e., the causal plane $F \times H$.
Our estimations using information quantifiers in the causal plane $C \times F$ (or causal plane $H \times F$) are in agreement with the idea that Fisher information saturates as the size of the system grows. Overall our findings show that Fisher information reach a maximum as the number of interconnected neurons grows.

But if we considered the case of uncorrelated neurons, then Fisher information would increase linearly with the number of neurons in the population \cite{Seung1993}. When accounting the efficiency of coding
information in the second order statistics of the population responses, it has been theoretically argued
that Fisher information of this system would grow linearly with its size \cite{Shamir2001}.
Figure~\ref{figu0} A, C, and E show that the variance grows as the number of interconnected neurons becomes higher.
Thus the Eq.~(\ref{eq:Fhis}) implies that Fisher information increases as well, considering the encoding variable is strictly larger than two \cite{Sejnowski99}. This would leads
to a very odd and unexpected neurophysiological behavior as when evaluating the computational capacity of
cortical population in sensory areas the total information should saturate at a level below the amount
of information available in the input \cite{Averbeck2006}. Importantly, as positive correlations vary smoothly with
space they drastically suppress the information in the mean responses,
and Fisher information of the system should saturate to a finite value as
the system size grows \cite{Shamir2001,Averbeck2006}.

We showed that accounting for the ordinal structures present in the ISIs and their local influence on the associated probability distribution allows us to estimate how information saturates when the number of interconnected neurons increases. These findings are in agreement with the hypothesis of Shamir \cite{Shamir2001} and Averbeck \cite{Averbeck2006}.
That is, as the number of interconnected neurons increases, correlations must be such that information
reaches a maximum \cite{Shamir2001,Averbeck2006,Montani2013}.
The amount of information that can be processed by the nervous system is finite,
and it cannot be larger than
the amount of extracted information. As data become available for a very large number of neurons,
our theoretical approach could provide an important mathematical tool to investigate
how information of the system saturates to a finite value as
the system size grows.
Further investigations should exhaustively account for the problem of finite size effects in the lattice and different external stimuli. This will allow us to investigate how
quickly information saturates as the number of neurons increases,
and whether it saturates or not at a level well below the amount
of information available in the input.

Understanding of the mechanisms of cortical processing of sensory
information requires the investigation of the dynamics
between different excitatory and inhibitory cortical circuitry \cite{Xu2009}.
When large numbers of inhibitory neurons are considered, typically about 40\% of the population, they are likely
to have differing roles in inhibiting pain or itch \cite{Polgar2013}.
Here, we use an alternative methodology when considering a large population of neurons, we showed that by building the phase space of Fisher information versus
MPR statistical complexity/permutation entropy it is possible to characterize the dynamics of the system, and to find the optimal values that would maximize information transmission within a neuronal network.
Our approach could be an useful tool to investigate
how the loss of inhibitory neurons marks the dynamics of autism in mouse, and to study
the essential role that inhibitory neurons could play on the development of proper circuitry in the cortex \cite{Meechan2009,
Gonchar2007, Just2004, Dolen2007, Gonchar2008}. Moreover it can be of help to examine neuronal avalanches as the balance between excitation and inhibition controls its temporal dynamics \cite{Lombardi2012}.

Shannon developed the Information Theory, as a measure of the uncertainty in a message
while essentially inventing what became known as the dominant form
of ``information theory'' \cite{Shannon}. As we consider a time series  ${\mathcal X}(t) \equiv \{ x_t ; t = 1, \cdots , D \}$,
if $x_t$ attain a finite number $D$ of values, the classical Shannon entropy measures the mean conditional uncertainty of the future $x_{t+1}$ given the whole past $x_{t-D}, \ldots , x_t$. Thus, we have $0 \leq S  \leq \ln (D)$, with $S = 0$ if the series is perfectly predictable from the past and $S = \ln (D)$ if and only if all values are independent and uniformly distributed.
Fisher information, $F[P]$, constitutes a very useful local measure to detect changes in the dynamic behavior.
By applying causal Fisher information it is possible to quantify changes in the dynamics of a system, thanks to the sensitivity of this measure to alterations in the associated probability distribution.

The informational causal plane of Fisher information versus MPR statistical complexity, $C\times F$ (or Fisher versus Shannon entropy $H \times F$), quantifies the local versus global features of the dynamics in the system under study.
Consequently, when considering a row signal of spontaneous ISI activity,
building the causal plane $C \times F$ (or causal plane $H \times F$) provides us with a useful tool to detect and quantify the optimal biophysical values that allow the neural network to transmit information more efficiently.
On the other hand, the causal informational plane complexity versus entropy , $H \times C$, quantifies global versus global characteristics of the distribution
failing to prove, therefore, information of the local features of the causal information.

Regarding the potential limitations of the methodology, note that in order to avoid the bias deviation problem and to have reliable statistic it has been established that the condition $M >> D!$  must be satisfied \cite{Bandt2002, Rosso2007}. We have chosen an embedding delay $\tau = 1$, but other values of $\tau$ might also provide further information \cite{Zunino2010B,Soriano2011,Zunino2012}.
The Bandt and Pompe procedure does not specify which order for generating ordinal patterns should be privileged.
That is, how to specify the ordering of index series ${i}$ and, consequently, for a pattern length $D$,
we have $D!!$ possible $i$-sequences.
We face therefore an ambiguity when local information measures are evaluated. The issue does not affect at all the
evaluation of global entropic quantifiers (like Shannon entropy). Different orders would lead, though, to different
``local" information contents.
As in the case of the PDF-histogram, we can reduce drastically the number of available $D!!$ possibilities if we proceed
to form patterns of length $D$ starting from those of length $D - 1$.
However, some lack of precise definition remains in the assignation of the pattern indices ${i}$.
In our current paper, we used the lexicographic ordering given by the algorithm of Lehmer \cite{Lehmer}, amongst other possibilities, due to
it provide the better distinction of different dynamics in the Fisher vs Shannon plane (see \cite{Olivares2012A,Olivares2012B}).

\section{Conclusions and perspectives}
\label{Sec:Conclusions}

Recording the spiking activity of a very large number of neurons is extremely difficult, and no one has recorded yet the activity of 1000 (even less, 500) neurons simultaneously. Consequently, when considering a large number of neurons the computational models developed by Izhikevich are the relevant ones \cite{Izhikevich2003,Izhikevich2004,Izhikevich2006,Izhikevich2007}.
Our paper presents a methodology to characterize the efficiency of large ensembles of neurons. It is important to point out that the interplay between conduction delays and STDP helps the spiking neurons to produce stable firing patterns
that would not be possible without these assumptions. Each neuron in the network is described by the simple model of spiking neurons of \cite{Izhikevich2003}
which has been described above in Eq. (1) to Eq. (3). In this simple model we choose to account for the STDP, with large proportion of inhibitory neurons, as its interplay with conduction
delays helps the spiking neurons to spontaneously self-organize into groups with patterns of time-locked activity.
Thus, it produces stable firing patterns emulating spontaneous brain activity that would not be possible without this assumption.
We have shown that an increasing degree of interconnectivity $m$ implies greater data dispersion. In subsection 3.1 we have explicitly shown
that the variance does not reach a maximum for some given value of $m$ as the interconnectivity becomes higher.
Thus, this kind of classical analysis would suggests that information does not saturate as the degree of interconnectivy becomes higher. Eq.~(\ref{eq:Fhis}) implies that Fisher information increases as the network connectivity becomes higher when considering a large
population of neurons. As the estimations through the variance are not accounting for the ordinal
``structures" present in ISIs,  they are far from the expected neurophysiological behavior.
In other words, noise correlation causes the amount of information in a population of neurons to saturate as the number of interconnected neurons grows \cite{Shamir2001,Averbeck2006}.
Our approach allows us to capture such behavior by introducing a simple and robust method that takes into account the time causality of the ISIs.

In this paper, we show an application of complexity measures based on order statistics
to simulation results that resemble neuronal activity when considering a very large number of neurons.
This approach provides us detailed insights into the dynamics of the neuronal networks. The choices of the different lattice sizes of $N=800$, $N=900$ and $N=1000$ are made to resemble cortical hypercolumns having a percentage of about 40\% of inhibitory neurons. The dynamic of the network is then investigated choosing different degrees of network interconnectivity.

Summing up, the current approach provides us a novel methodology to investigate
the causal information structure of the ISIs when considering
a row signal of spontaneous activity of a very simple neuronal network.
We are currently working to include different network structures in the inputs
to account for external stimuli
within the simulated data, and to investigate
the finite size effects considering different grid sizes.
We also plan to extend the current approach to cases in which external stimuli such as
different discrete conductivity values are considered.
This would require extending the current approach currently designed for entropy like
quantities to entropy transfer quantities \cite{Schreiber2000}.
As non-causal mutual information fails to distinguish information that
is actually exchanged from shared information due to common history and input signals,
the current approach could be very useful when extended to transfer entropy-like quantities.
This is not only important from a theoretical point of view, it might help determine which areas of the cortex could have a higher level of information, and to evaluate how causal interactions in neural dynamics would be modulated by behavior.
We believe that this will become an important tool for future research on the encoding capacity of biologically realistic neural
networks.

\section*{Acknowledgments}
Research supported by PIP 0255/11 CONICET, Argentina (FM).
F. Montani and O. A. Rosso acknowledge support by CONICET, Argentina.
O. A. Rosso gratefully acknowledges support by FAPEAL, fellowship, Brazil.

\newpage

\end{document}